# Evidence for Magnetic Fields in Light Meson Spectra

**David Akers**
*Lockheed Martin Corporation, Dept. 6F2P, Bldg. 660, Mail Zone 6620,*
*1011 Lockheed Way, Palmdale, CA 93599, USA*

**Summary. --** Mac Gregor's constituent-quark model is reviewed with currently published data from light meson spectroscopy. It was previously shown that magnetic sources were responsible for the quantization of several mass-splittings in Mac Gregor's model. The existence of a 70-MeV quantum was postulated by Mac Gregor and was shown to fit the Nambu empirical mass formula $m_n = (n/2)137m_e$, $n$ a positive integer. It is shown in this paper that the light meson spectra also fit into the constituent-quark model and are in agreement with the Russell-Saunders coupling scheme. The existence of magnetic fields is suggested by the successful accounting of these meson spectra.

## 1. -- Introduction.

In the present paper, we review the role that Mac Gregor's constituent-quark model plays in determining particle mass [1, 2]. A brief review of the model is given in sect. **2**. In sect. **3**, the Russell-Saunders coupling scheme is restated in terms of two inequivalent quarks, forming the bound states of mesons. We identify the triplet *P*-states for mesons which can be split into separate energy levels. The experimental evidence for Russell-Saunders splittings of meson states is presented in sect. **4**. In sect. **5**, we discuss the theoretical underpinnings of the experimental data in terms of the existence of magnetic fields generating the Russell-Saunders splittings. Finally, concluding remarks are made in sect. **6**.

**2. -- Mac Gregor's Constituent-Quark Model.**

In the course of researching particle physics over several decades, Mac Gregor [1] developed a constituent-quark (CQ) model of elementary particles and showed that the CQ masses were directly related to the masses of the electron, muon, and pion. Later, a connection was discovered by Akers [2] between Mac Gregor's model and Nambu's empirical mass formula $m_n = (n/2)137m_e$, $n$ a positive integer and $m_e$ the mass of the electron. A comparison of Nambu's empirical mass formula with Mac Gregor's CQ model can be found in [2]. Mac Gregor's esoteric notation included a 70 MeV quantum, a boson excitation B with the mass of the pion at 140 MeV, a fermion excitation F with a mass of 210 MeV or twice the muon mass, and a 420 MeV excitation quantum X. The 70 MeV quantum and the 420 MeV quantum X do not correspond to any observed particles but serve as the building blocks of mesons and baryons in the CQ model. The mechanism for generating the CQ masses is discussed in Mac Gregor [1] and will not be repeated here. It is sufficient to say that we shall accept the evidence of Mac Gregor's CQ model from its previous agreement with experimental data [1, 2] and that we shall present further evidence from the Particle Data Group [3]. This evidence is presented in sect. **4** after we combine the CQ model with the Russell-Saunders coupling scheme.

**3. -- Russell-Saunders Coupling.**

Mesons are composites of two quarks or quark-antiquark pairs and have been extensively studied as quarkonium and compared to nonrelativistic positronium-like bound states, which are observable as narrow resonances in electron-positron annihilation. If these quark-antiquark states are indeed comparable to atomic-like systems, then it would seem logical that the physics of the atomic scale would be applicable to the particle scale. In this section, we recall the Russell-Saunders coupling scheme [4] of two inequivalent particles and apply this scheme to the quarks in light meson spectroscopy.

The Russell-Saunders coupling scheme assumes that the electrostatic interaction in atomic systems between two inequivalent electrons dominates the spin-orbit interaction. The orbital momenta and the spins of the particles couple separately to form $\boldsymbol{L} = \boldsymbol{L}_1 + \boldsymbol{L}_2$ and $\boldsymbol{S} = \boldsymbol{S}_1 + \boldsymbol{S}_2$. Then the total angular momentum is given by $\boldsymbol{J} = \boldsymbol{L} + \boldsymbol{S}$. For each $l$ and $s$, the $j$ values are $|l + s|, \ldots, |l - s|$. The combinations are shown in table I.

**Table I.** Russell-Saunders Coupling of Two Inequivalent Particles

| $l$ | $s$ | $j$ | Spectral Terms | Number of States in a Magnetic Field (Number of $m_j$ Values) |
|---|---|---|---|---|
| 2 | 1 | 3, 2, 1 | $^3D_{1, 2, 3}$ | 3 + 5 + 7 = 15 |
| 2 | 0 | 2 | $^1D_2$ | 5 |
| 1 | 1 | 2, 1, 0 | $^3P_{0, 1, 2}$ | 1 + 3 + 5 = 9 |
| 1 | 0 | 1 | $^1P_1$ | 3 |
| 0 | 1 | 1 | $^3S_1$ | 3 |
| 0 | 0 | 0 | $^1S_0$ | 1 |
| | | ----------- | | ------------- |
| | | 10 States | | 36 States |

In table I, the number of states or number of $m_j$ values are shown. For the $^3P_{0, 1, 2}$ states, we have included the $(2j + 1)$ values or 1, 3, 5 for $j = 0$, 1, and 2. If mesons are composites of two quarks, then we have two inequivalent quarks by reason of charge conjugation alone or by reason of flavor type in some cases. Therefore, we would expect that the number of triplet $P$-states in meson systems to reflect the order of table I with the

$^3P_1$ states splitting into 3 separate levels and with the $^3P_2$ states splitting into 5 separate levels in the presence of a magnetic field. In the next section, we examine the experimental evidence for Russell-Saunders splittings in light meson systems.

**4. -- Light Meson Spectra.**

As noted in sect. **2**, we utilized a scale of particle masses based upon the CQ model as found in fig. 3 of the paper by Mac Gregor [1]. In fact, there are two distinct scales in Mac Gregor's fig. 3; one scale starts with the pion mass at 140 MeV and has steps of X = 420 MeV, and the other scale starts at zero and has steps of q = 315 MeV. The X = 420 MeV scale has particle masses at π(140), η(547), η'(958), η(1440), η(1760), and η(2225). The q = 315 MeV scale has particle masses at η(1295), η(1580), and D(1864); these particles can be identified as composites of 4q, 5q, and 6q masses, respectively. Binding energies are also discussed in Mac Gregor's work [1].

The first set of particles corresponding to the 420 MeV scale are shown in fig. 1. In fig. 1, the experimental meson masses are indicated by solid lines and are taken from the Particle Data Group [3]. The vertical arrows represent energy separation of about 420 MeV between states. We note a consistent pattern of energy separation between the spin-singlet and -triplet states. The lowest lying charmonium states are also shown for comparison, and a 420 MeV energy separation is shown by the indicated arrows. A missing $f_0$ meson is shown at 560 MeV and is predicted to exist.

In fig. 2, it is shown that there is a consistent pattern of spin-spin and spin-orbit energy separation between the states. The set of particles η'(958), ϕ(1020), $f_0$(1370), $f_1$(1465), and $f_2$(1525) parallel the lowest lying charmonium states. This pattern is repeated with the set of particles η(1295), ω(1420), $f_0$(1710), $f_1$(1805), and $f_2$(1850). The low meson masses $f_1(1465)$ and $f_1(1805)$ are predicted to exist. The pattern of energy separation is repeated with additional sets of particles as shown in figures 3 and 4. In these figures, the solid lines represent experimental meson masses, and the dashed-dot lines represent unobserved mesons, which are predicted to exist.

An additional set of mirror states to the charmonium states is shown in fig. 5. These are isospin $I = 0$ mesons, and they have a symmetry about the particle mass of 2397 MeV, which is the classical Dirac monopole mass. For a discussion of these states see the paper by Akers [5]. We note that the spin-spin and spin-orbit energies are similar in separation between their respective states.

Starting with the first set of particles associated with η'(958) in fig. 2, we search for particle states, which satisfy number of states for Russell-Saunders coupling in table I. The results of this search are shown in fig. 6. In fig. 6, there are indications of particle states, which fit the Russell-Saunders coupling scheme with 1, 3, and 5 states for the triplet *P*-states ($l = 1$, $s = 1$). With the introduction of $a_2(1320)$ into these states, we have evidence for 5 mass splittings in the $^3P_2$ states consistent with the Russell-Saunders coupling scheme. Experimental masses are indicated with solid lines. Those indicated by dashed-dot lines are predicted to exist.

We can repeat the process of searching for particle states, which fit the Russell-Saunders coupling scheme by utilizing the scale determined from the CQ model. These results are shown in figures 7 to 14. A total of 9 figures, including fig. 6, represent evidence consistent with the Russell-Saunders coupling scheme of table I.

## 5. – Discussion.

It was noted in the last section that the evidence for Russell-Saunders or *LS* coupling is derived from the light meson spectra of figures 6 to 14. If these spectra can be shown to satisfy the Lande interval rule, which is widely used in atomic, molecular and nuclear physics, then we have evidence for the presence of Russell-Saunders or *LS* coupling. The Lande interval rule is the test. For the *P*-states, the Lande interval rule predicts a mass splitting or ratio of 2.0 for the states in the same multiplet. As shown in table II, the Lande interval rule is satisfied.

**Table II.** Calculations of the mass splittings for *P*-states ($J = 2$) associated with the mesons η'(958), $f_0$ (1500), η(1295), η(1440), and η(1580).

| | |
|---|---|
| η(1295) | Upper states: [X(1850) – $a_2$(1750)]/[$f_2$(1810) – $a_2$(1750)] = 1.7 |
| | Lower states: [$a_2$(1750) – $f_2$(1565)]/[$a_2$(1750) – $f_2$(1640)] = 1.7 |
| η(1440) | Upper states: [$f_2$(2010) – X(1900)]/[$f_2$(1950) – X(1900)] = 2.2 |
| | Lower states: [X(1900) – $a_2$(1750)]/[X(1900) – $f_2$(1810)] = 1.7 |
| η'(958) | Upper states: [$f_2$(1525) – $f_2$(1380)]/[$f_2$(1430) – $f_2$(1380)] = 2.9 |
| | Lower states: [$f_2$(1380) – $f_2$(1270)]/[$f_2$(1380) – $a_2$(1320)] = 1.8 |
| $f_0$(1500) | Upper states: [$f_2$(1640) – $f_2$(1525)]/[$f_2$(1565) – $f_2$(1525)] = 2.9 |
| | Lower states: [$f_2$(1525) – $f_2$(1380)]/[$f_2$(1525) – $f_2$(1430)] = 1.5 |
| η(1580) | Upper states: [$f_2$(2150) – $f_2$(2010)]/[X(2070) – $f_2$(2010)] = 2.3 |
| | Lower states: [$f_2$(2010) – X(1900)]/[$f_2$(2010) – $f_2$(1950)] = 1.8 |

The evidence of Russell-Saunders coupling in particle physics is established, and yet we are left with the problem of the source for the mechanism that generates the magnetic fields in the meson spectra. The existence of Russell-Saunders coupling, which is lifted in figures 6 to 14, is evidence for strong internal magnetic fields within the composite mesons. The effect of these strong internal magnetic fields is shown in figures 6 to 14; however, what is the cause of these magnetic fields within the particles? What are the sources of these fields? There are a number of possible sources for magnetic fields: 1) magnetic charges, 2) magnetic dipole moments of the individual quarks, and/or 3) Ampere's Hypothesis – the assumption that all magnetism comes from electrical currents. There is strong evidence in particle physics that supports the second; namely, quarks possess magnetic dipole moments, which give rise to magnetic interactions, and particle mass.

## 6. – Conclusion.

In this paper, we reviewed the constituent-quark model of Mac Gregor [1], derived two particle mass scales to identify meson spectra, and presented experimental data derived from the Particle Data Group [3]. We presented experimental evidence for Russell-Saunders splitting of meson states, which has never been seen before now in particle physics. This evidence was tested by the Lande interval rule. The tabulated data satisfied the Lande interval rule.

It was suggested that strong internal magnetic fields of the composite mesons are responsible for the mass splittings of figures 6 to 14. These internal magnetic fields produce a Zeeman or Paschen-Back Effect that lifts the degeneracy of the *P*-states and generates distinct particles. Thus, we are left with light meson spectra which fit the Russell-Saunders coupling scheme and which satisfy the Lande interval rule.

## REFERENCES

[1] M. H. Mac Gregor: *Nuovo Cimento A*, **103** (1990) 983.
[2] D. Akers: *International Journal of Theoretical Physics*, **33** (1994) 1817.
[3] Particle Data Group: *European Physical Journal C*, **15** (2000) 1.
[4] E. E. Anderson: *Modern Physics and Quantum Mechanics* (W. B. Saunders Company, Philadelphia, PA) 1971, p. 312.
[5] D. Akers: *International Journal of Theoretical Physics*, **29** (1990) 1091.

**Figure Captions**

**Fig. 1.** Experimental meson masses, indicated as solid lines, are taken from the Particle Data Group [3]. The vertical arrows indicate a separation energy of about 420 MeV. The lowest lying charmonium states are shown. Note the consistent pattern of the separation energy between the spin-singlet and -triplet states. The $f_0(560)$ is predicted to exist.

**Fig. 2.** Low meson masses are shown for comparison to the lowest lying charmonium states. Note the consistent pattern of spin-spin and spin-orbit energy separations between the states. The low meson masses $f_1(1465)$ and $f_1(1805)$ are predicted to exist. The $f_2(1850)$ meson, as indicated by the doted line, has some evidence for its existence.

**Fig. 3.** Another set of low meson masses is shown in comparison to the charmonium states. Note the consistent pattern of energy separations between the groups. The X(1880) state is predicted to exist.

**Fig. 4.** Another set of low meson masses is shown in comparison to the lowest lying charmonium states. Note the consistent pattern of energy separations between the groups. The $D_0(2320)$ is predicted to exist.

**Fig. 5.** Zeeman splitting of isospin $I = 0$ mesons. The dashed line represents the mass of the Dirac monopole or 2397 MeV from classical calculations. The experimental masses are indicated by solid lines. The η meson at 1820 MeV is predicted to exist. Note the symmetry of the low mass mesons with the charmonium states about the dashed line.

**Fig. 6.** The set of low meson masses connected with the η'(958) state. Experimental meson masses are indicated with solid lines. Those indicated by dashed-dot lines are predicted to exist and fit the Russell-Saunders coupling scheme.

**Fig. 7.** The set of low meson masses without connection to a singlet state. Experimental meson masses are indicated with solid lines. Those indicated by dashed-dot lines are predicated to exist and also fit the Russell-Saunders coupling scheme.

**Fig. 8.** The set of low meson masses connected with the η(1295) state. Solid lines represent experimental meson masses taken from the Particle Data Group [3]. Dashed-dot lines are mesons which are predicted to exist. Note the pattern of mass splittings which fit the Russell-Saunders coupling scheme.

**Fig. 9.** The set of low meson masses associated with the η(1440) state. Solid lines indicate experimental masses taken from the Particle Data Group [3]. Dashed-dot lines represent mesons which are predicted to exist.

**Fig. 10.** The set of low meson masses connected with the η(1580) state, which is predicted to exist. Experimental meson masses are indicated with solid lines. Those indicated by dashed-dot lines are predicted to exist. The $^3P_2$ states fit the Russell-Saunders coupling scheme.

**Fig. 11.** The set of low meson masses associated with the $f_0(2100)$ state. The notation is the same as in Fig. 10.

**Fig. 12.** The set of low meson masses associated with the $\eta(1760)$ state. The notation is the same as in Fig. 10.

**Fig. 13.** The set of low meson masses connected with the D(1864) state. The notation is the same as in Fig. 10.

**Fig. 14.** The set of low meson masses associated with the $\eta(2225)$ state. The notation is the same as in Fig. 10.

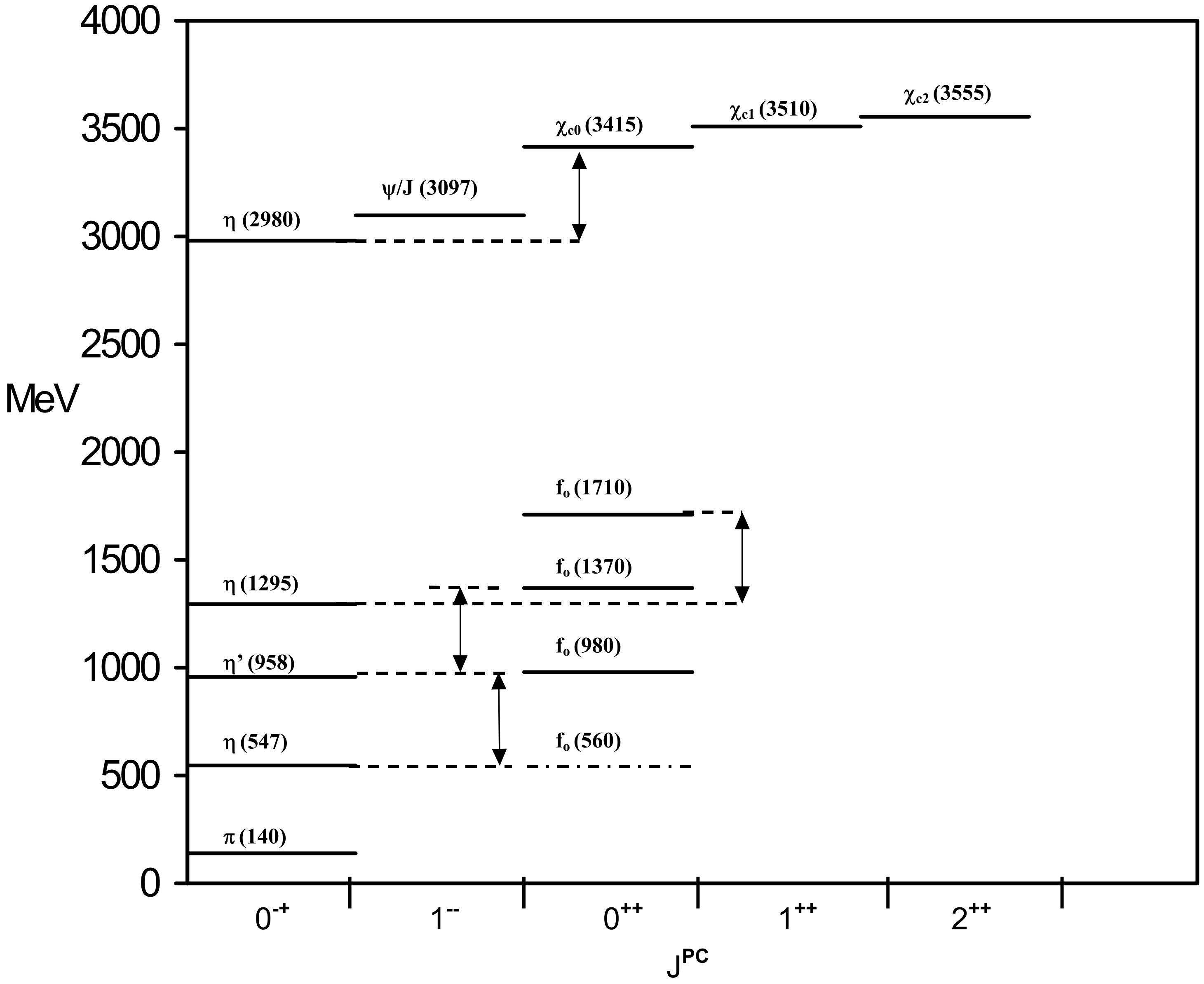

Fig. 1.

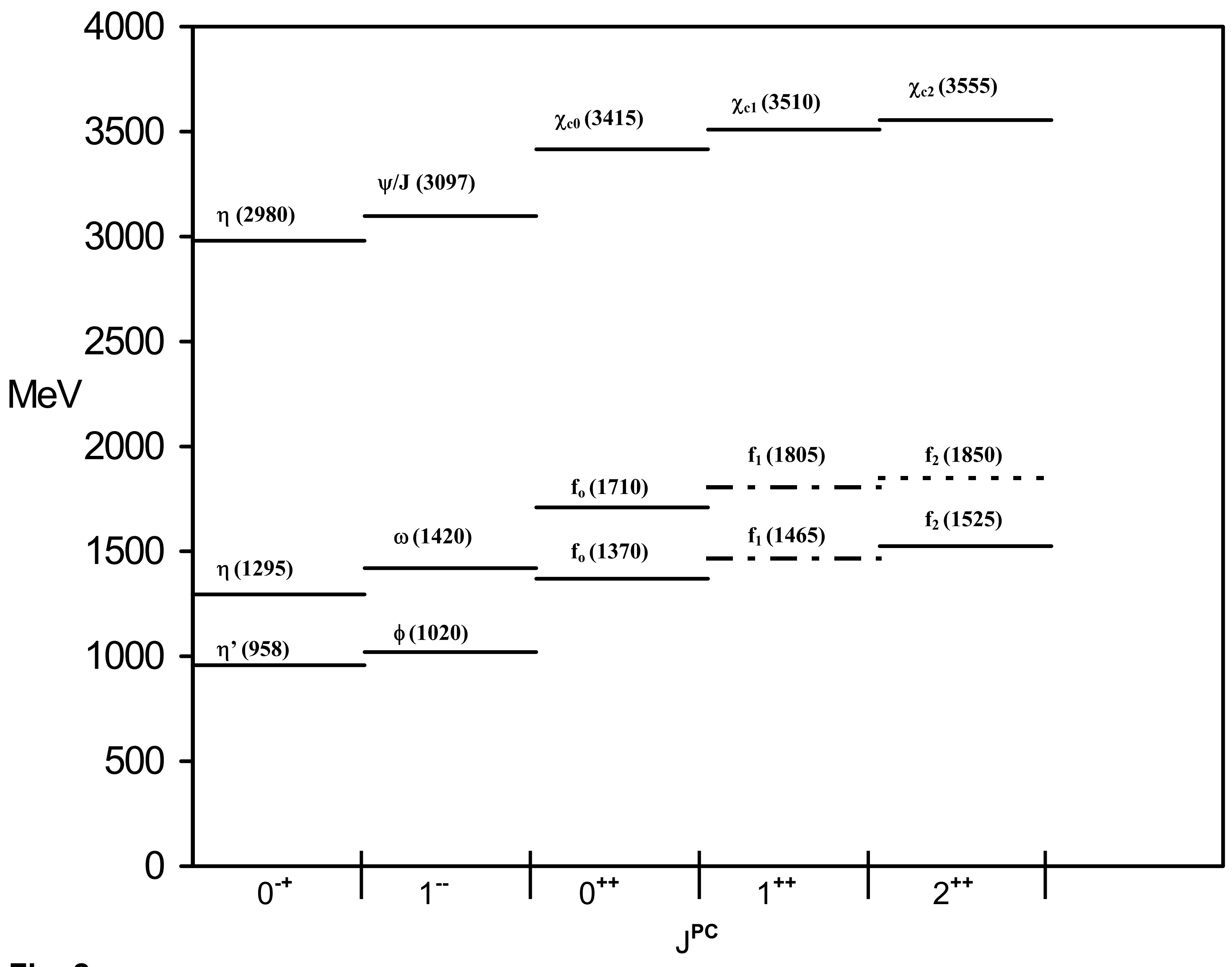

**Fig. 2.**

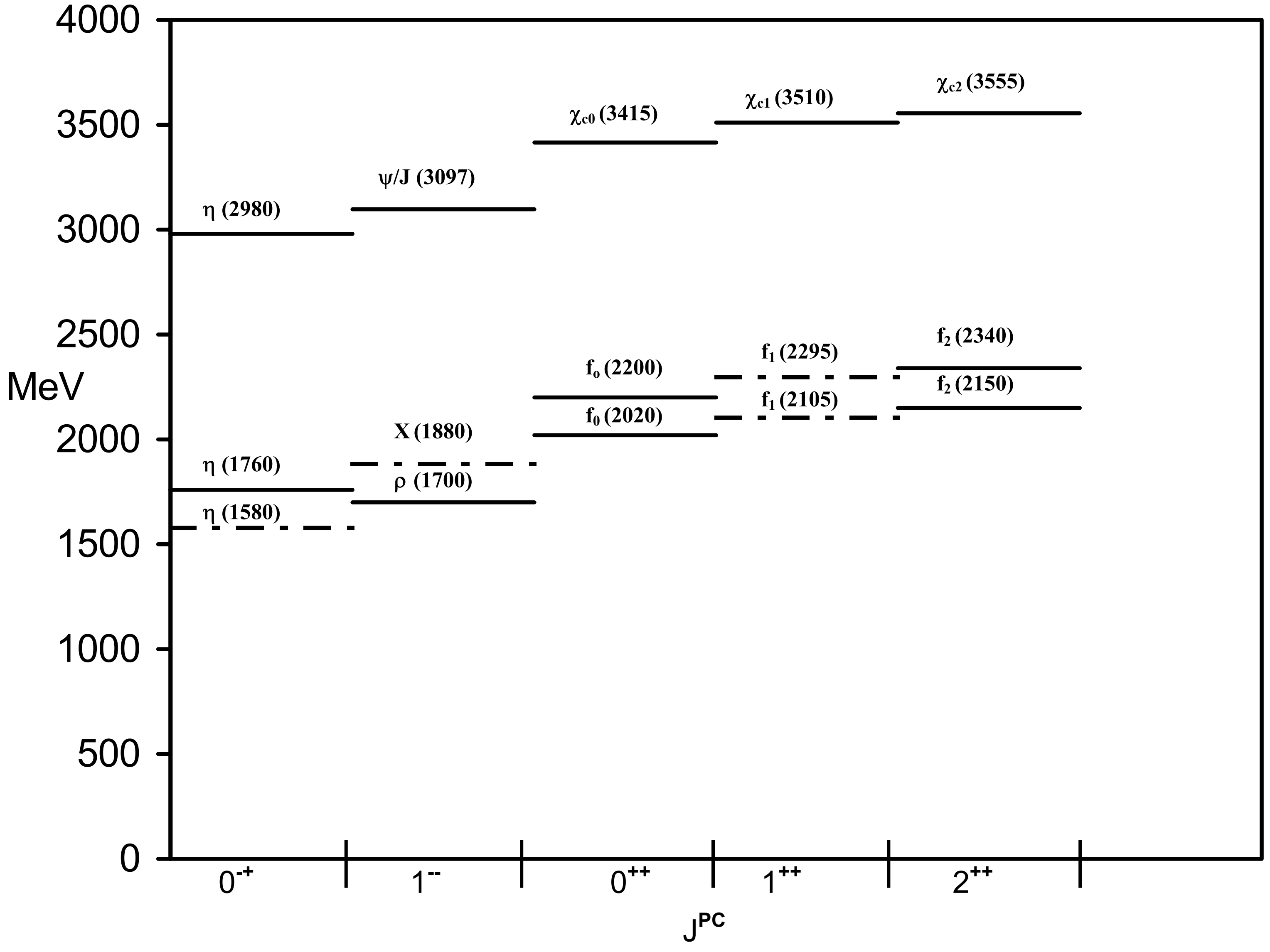

**Fig. 3.**

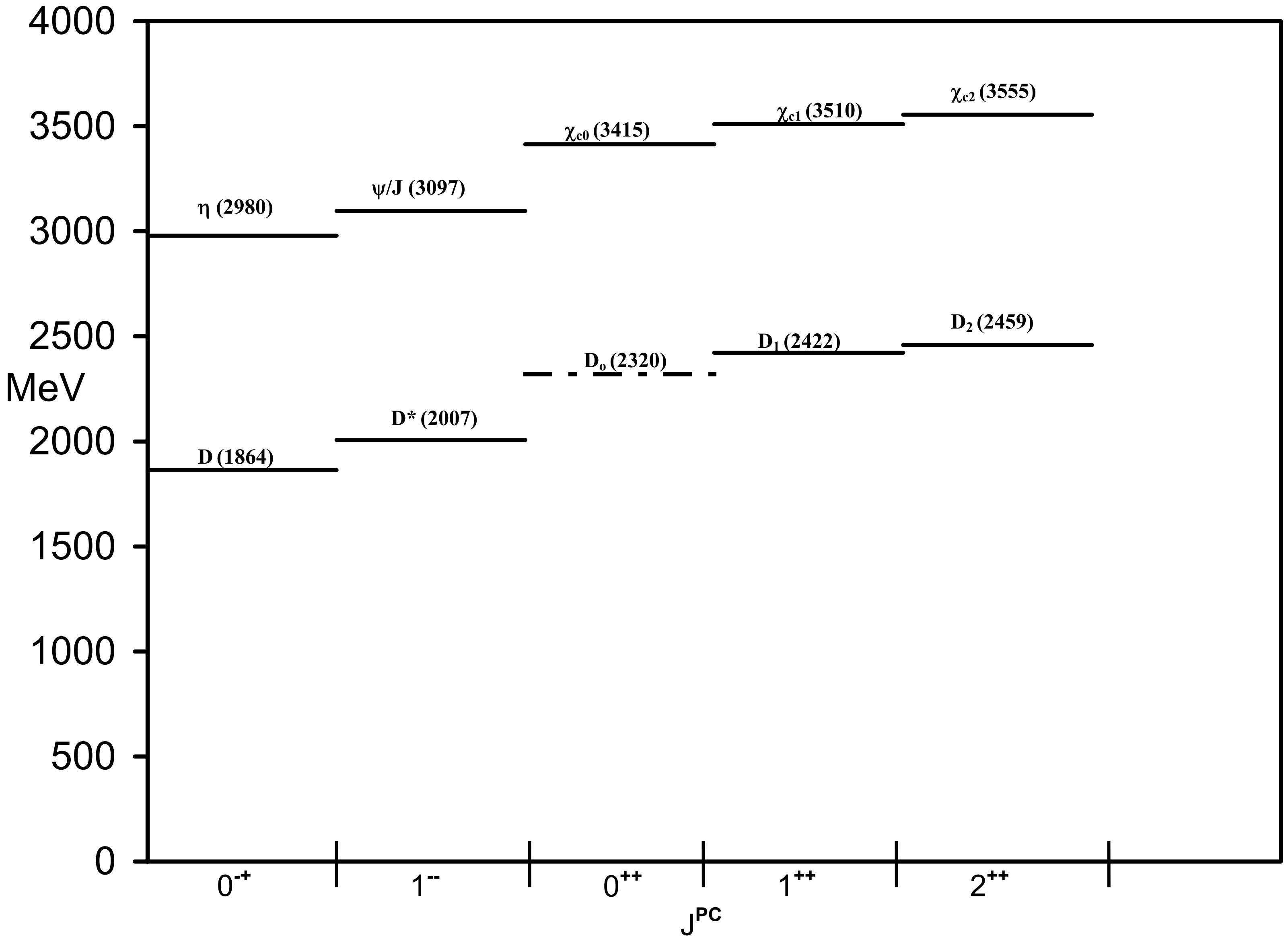

**Fig. 4.**

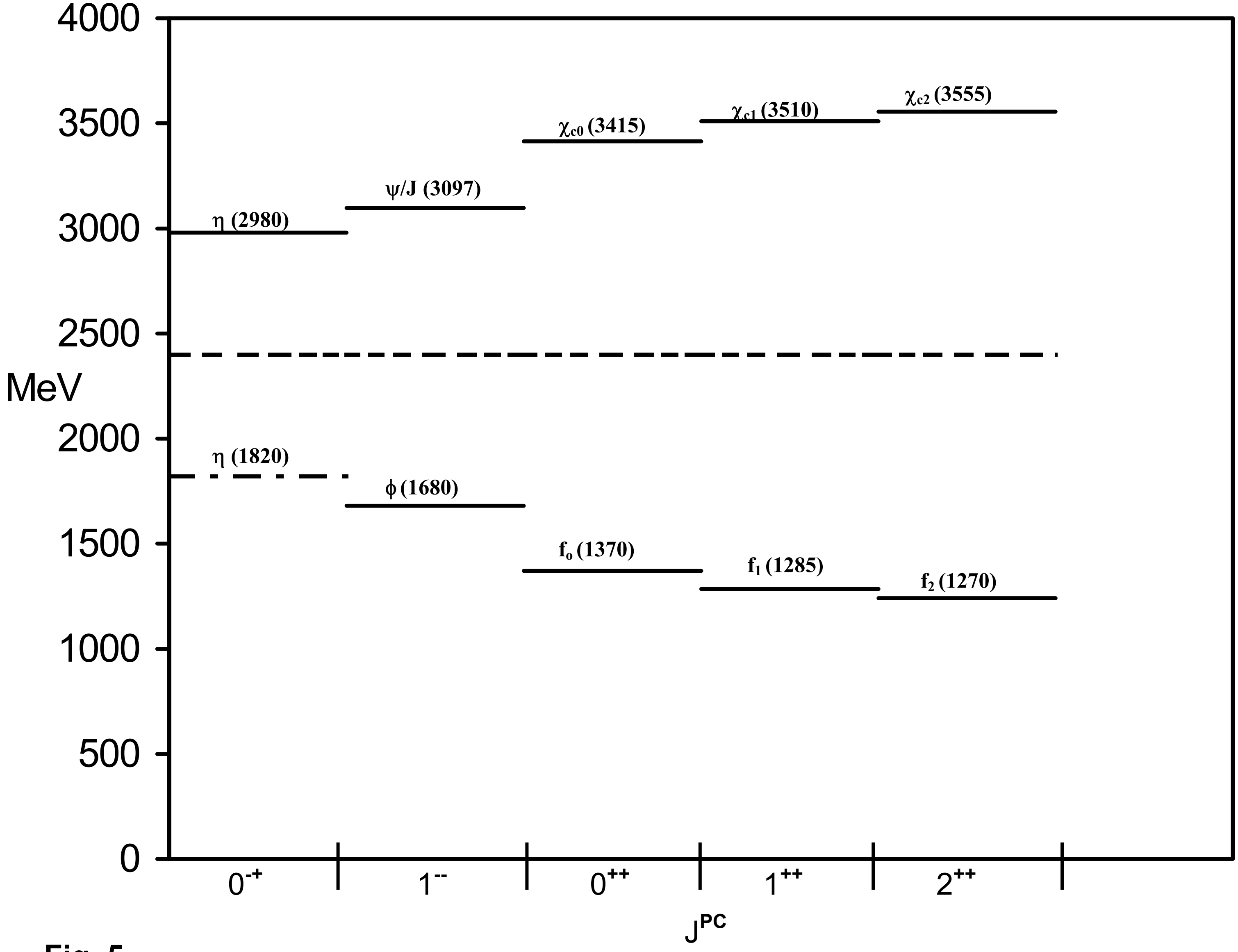

Fig. 5.

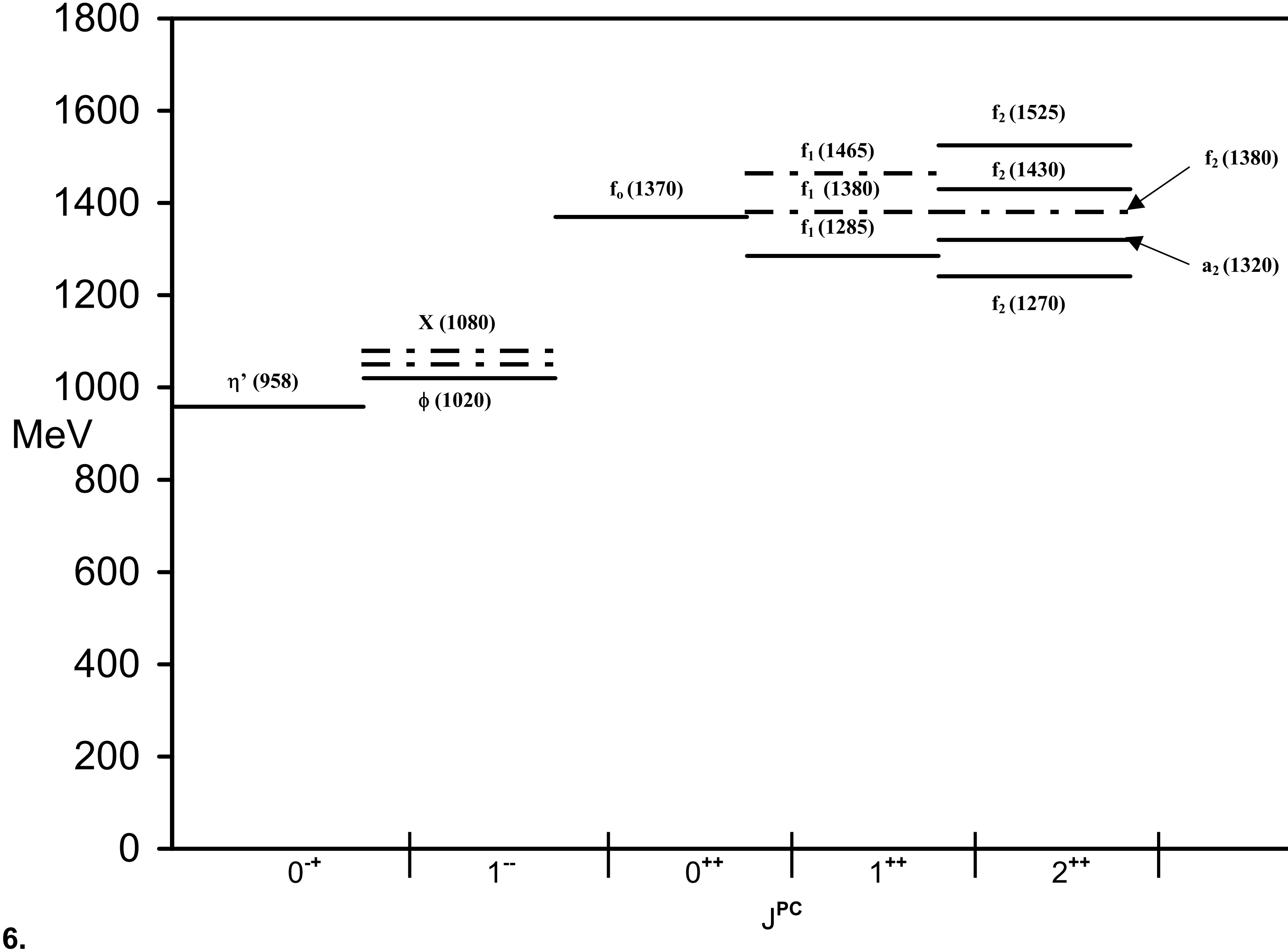

**Fig. 6.**

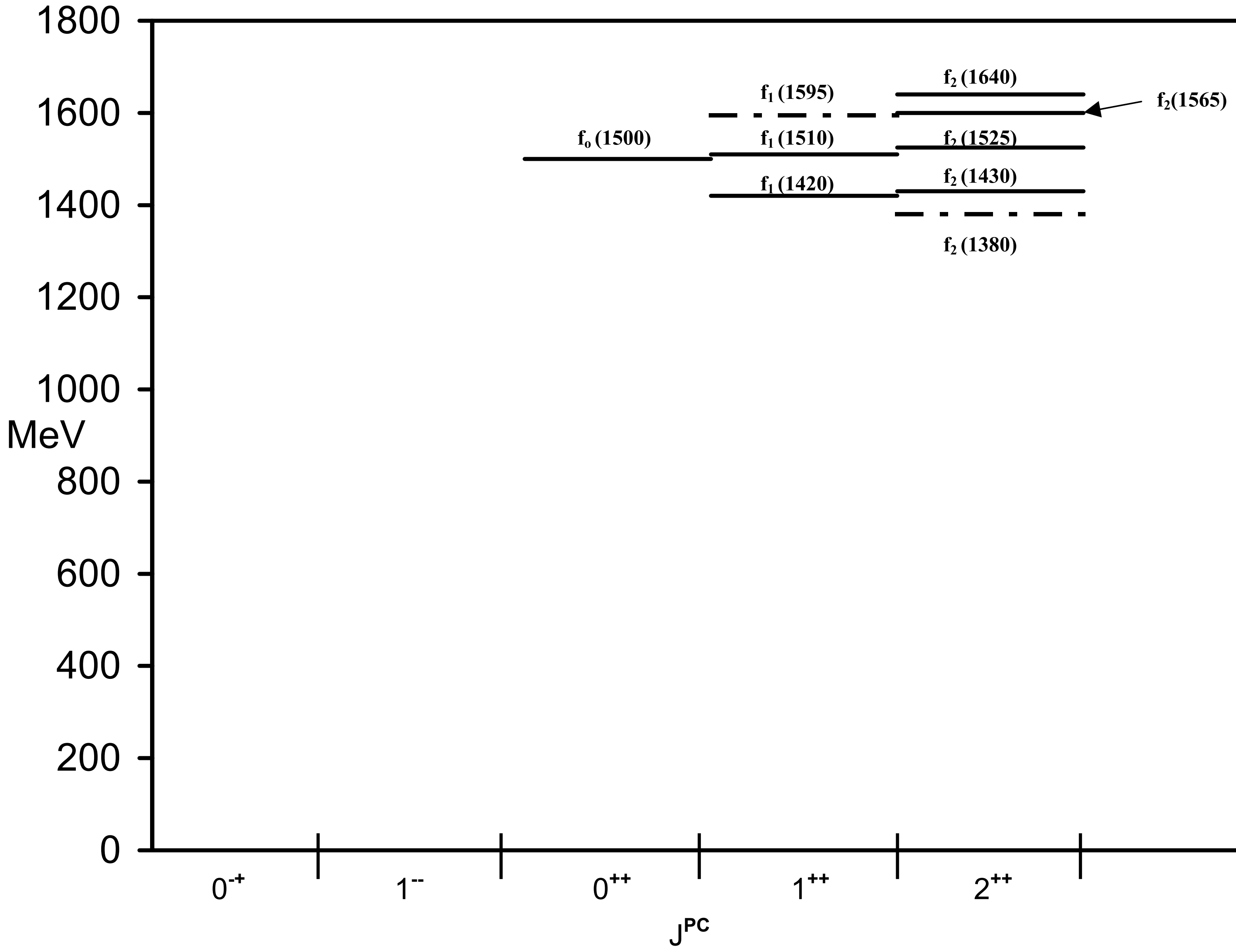

**Fig. 7.**

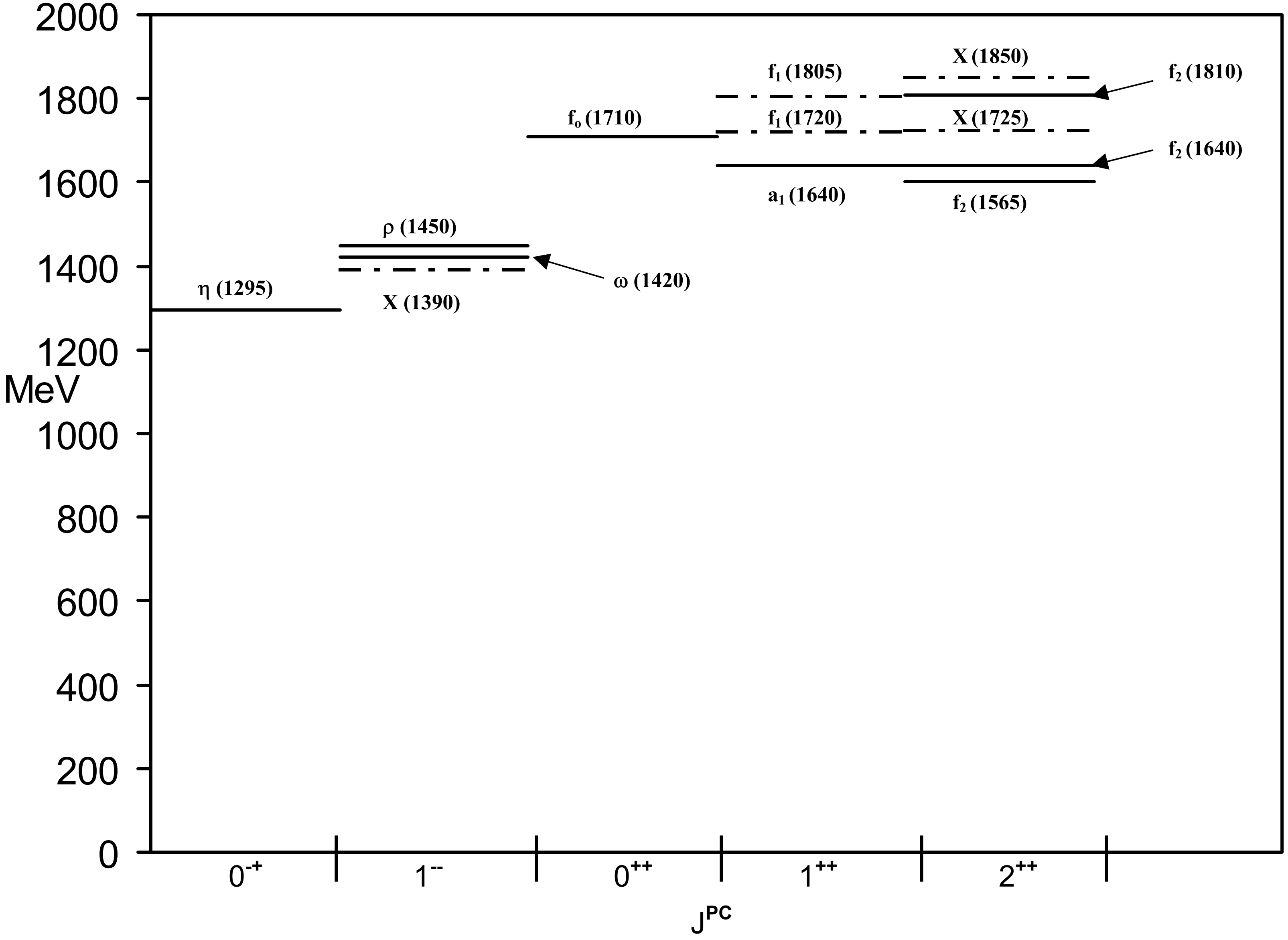

Fig. 8.

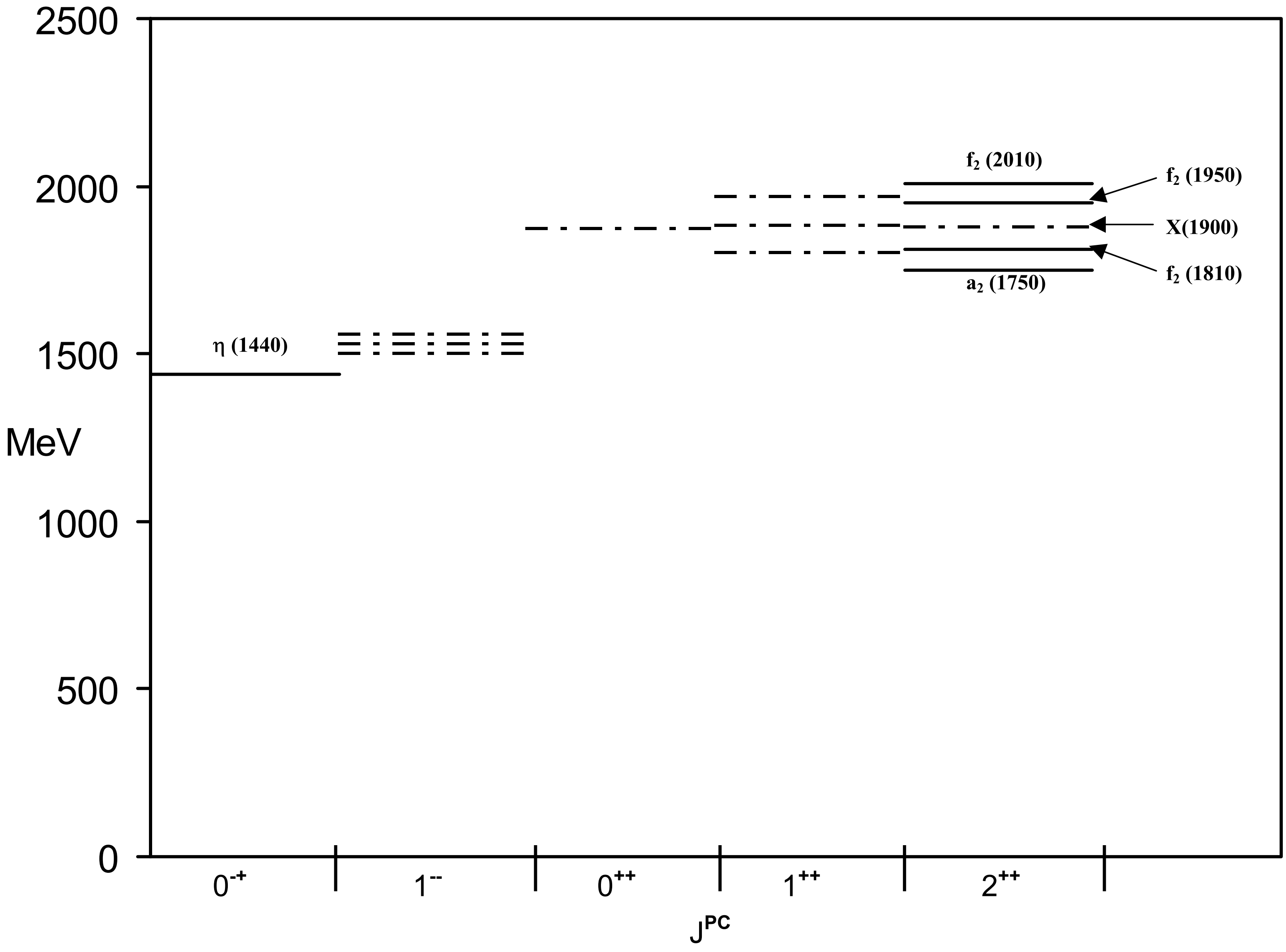

**Fig. 9.**

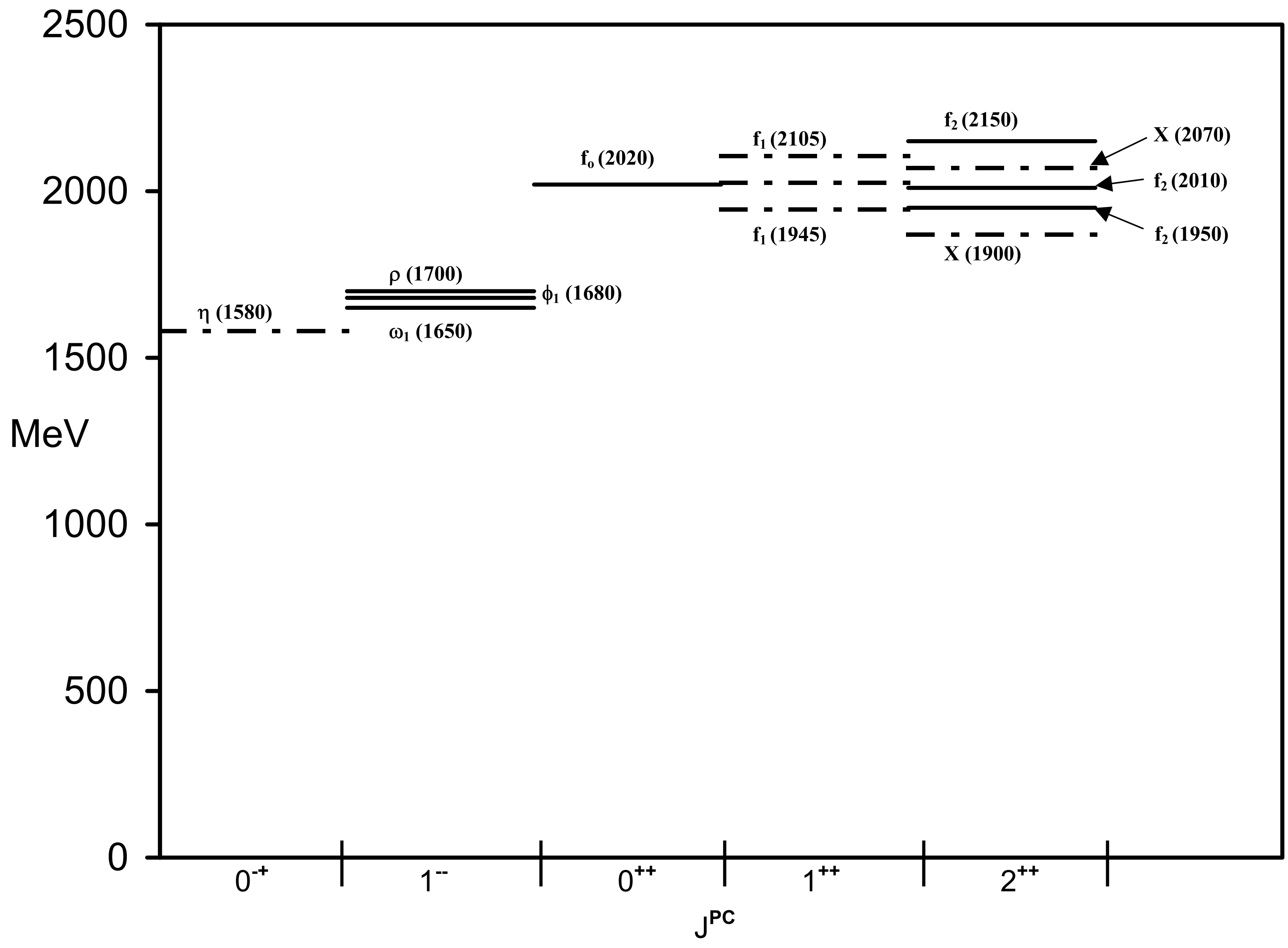

**Fig. 10.**

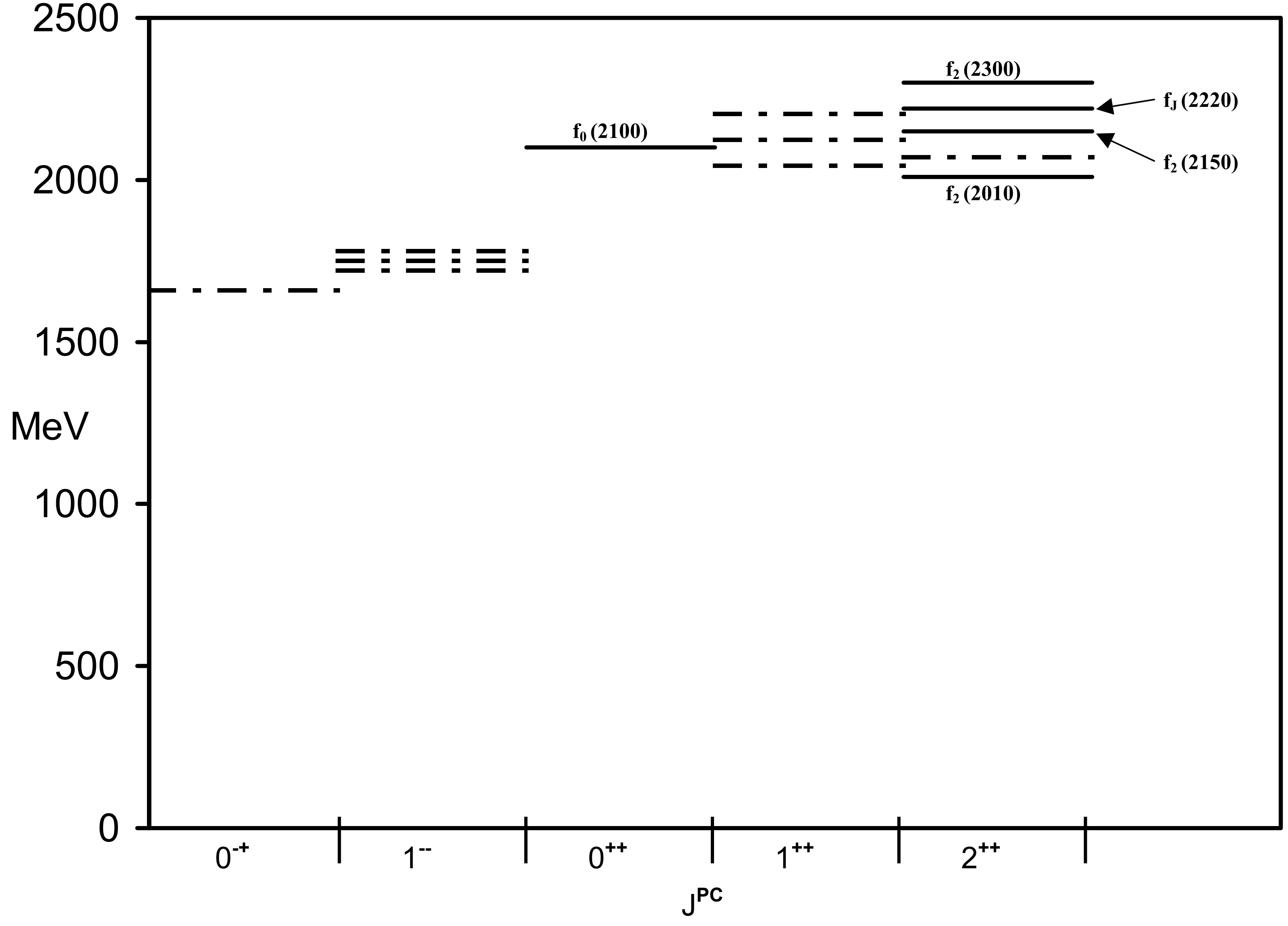

**Fig. 11.**

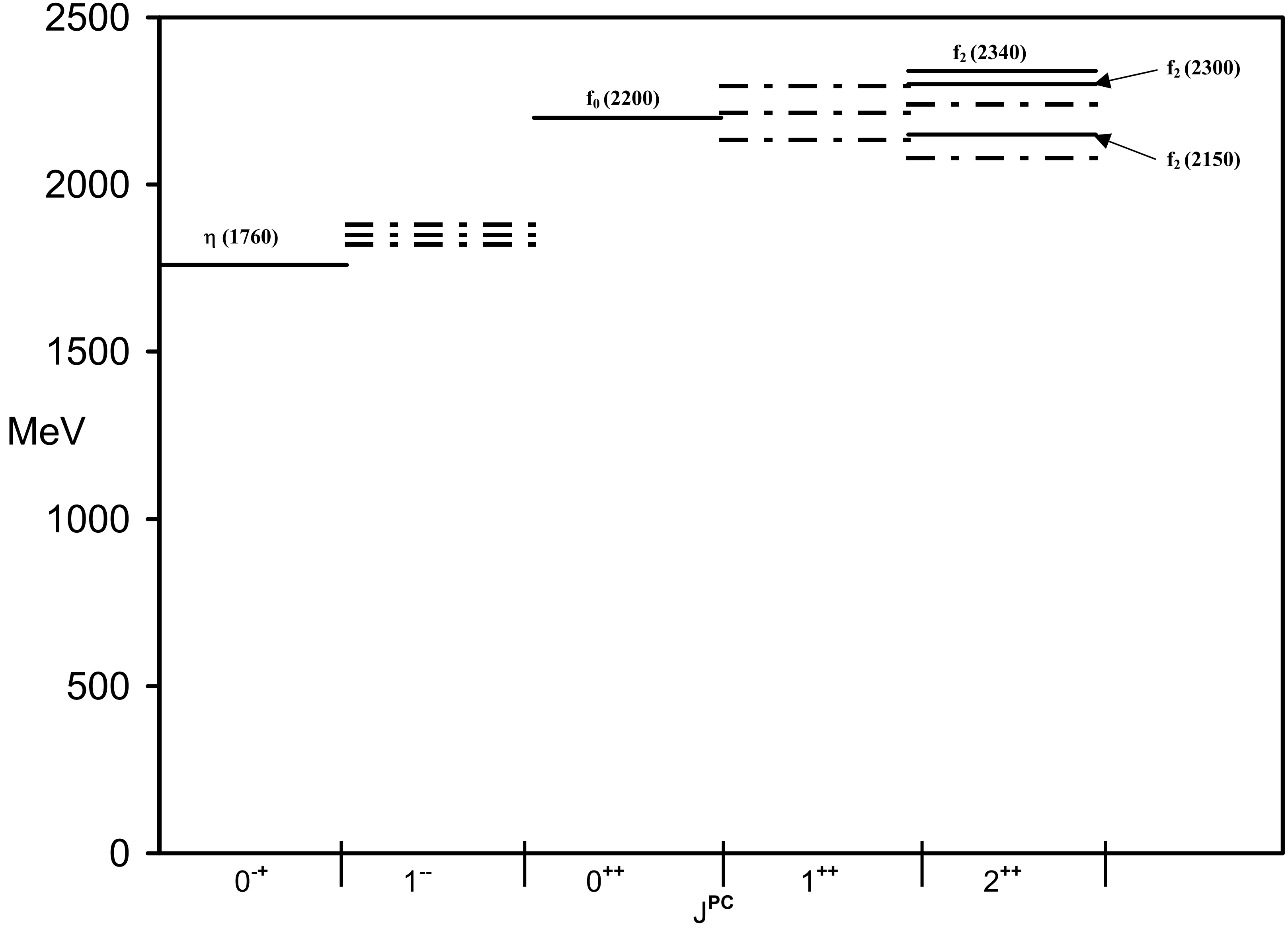

**Fig. 12.**

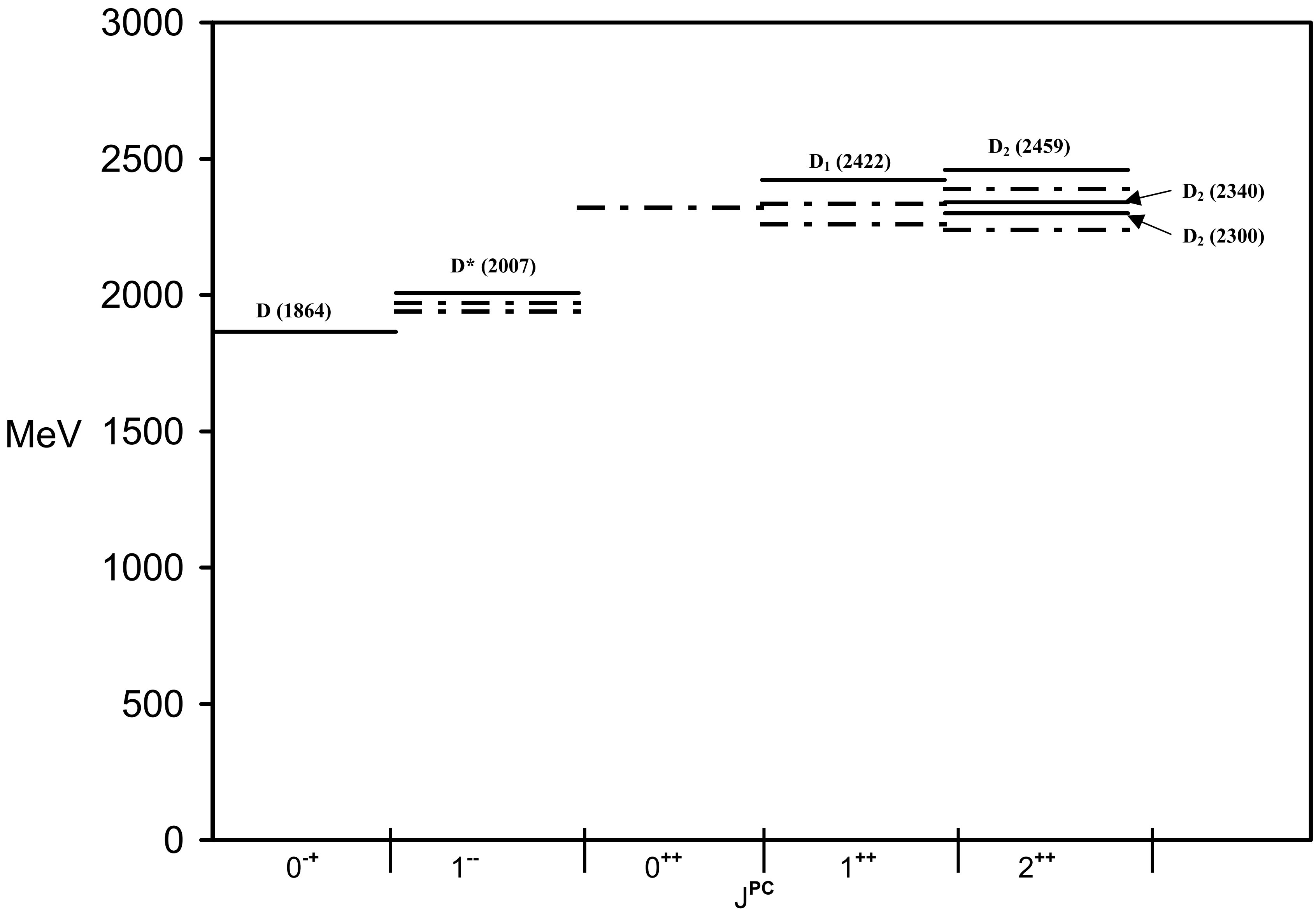

**Fig. 13.**

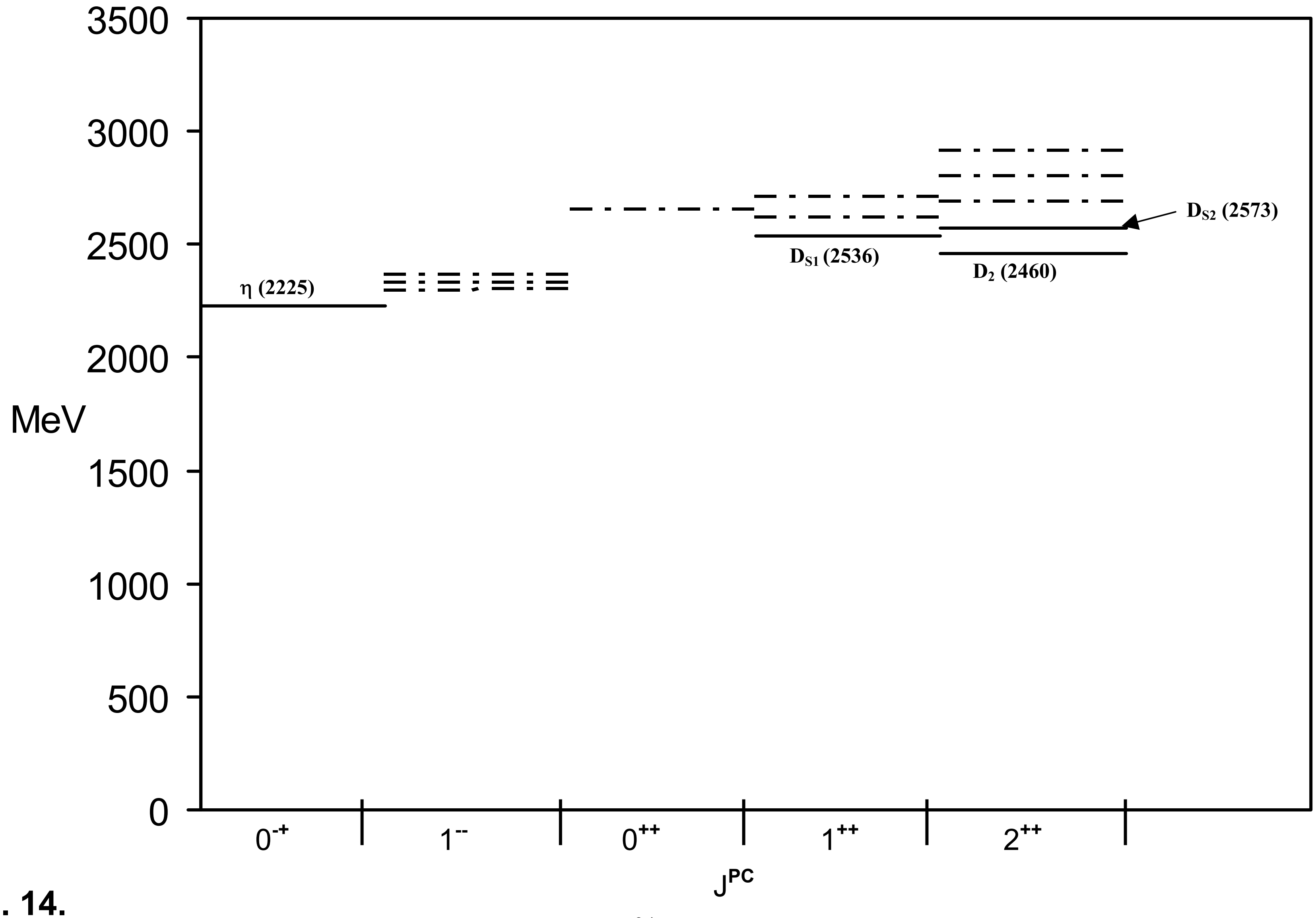

Fig. 14.